%% file: paper.tex
\documentclass[sigconf]{acmart}

\usepackage{multirow}
\newcommand\Tstrut{\rule{0pt}{2.4ex}}
\renewcommand{\arraystretch}{0.947}
\usepackage{subcaption}
\usepackage{listings}

\usepackage{tikz}
\newcommand\copyrightnotice[1]{
	\begin{tikzpicture}[remember picture,overlay]
		\node[anchor=south,yshift=20pt] at (current page.south) {\fbox{\parbox{\dimexpr\textwidth-\fboxsep-\fboxrule\relax}{#1}}};
	\end{tikzpicture}
}

\AtBeginDocument{%
  \providecommand\BibTeX{{%
    \normalfont B\kern-0.5em{\scshape i\kern-0.25em b}\kern-0.8em\TeX}}}

\copyrightyear{2022}
\acmYear{2022}
\setcopyright{acmlicensed}\acmConference[ARES 2022]{The 17th International Conference on Availability, Reliability and Security}{August 23--26, 2022}{Vienna, Austria}
\acmBooktitle{The 17th International Conference on Availability, Reliability and Security (ARES 2022), August 23--26, 2022, Vienna, Austria}
\acmPrice{15.00}
\acmDOI{10.1145/3538969.3538990}
\acmISBN{978-1-4503-9670-7/22/08}

\begin{document}

\title{Detecting Unknown DGAs without Context Information}

\author{Arthur Drichel}
\email{drichel@itsec.rwth-aachen.de}
\affiliation{%
	\institution{RWTH Aachen University}
	\city{}
	\country{}
}
\author{Justus von Brandt}
\email{justus.von.brandt@rwth-aachen.de}
\affiliation{
	\institution{RWTH Aachen University}
	\city{}
	\country{}
}
\author{Ulrike Meyer}
\email{meyer@itsec.rwth-aachen.de}
\affiliation{%
	\institution{RWTH Aachen University}
	\city{}
	\country{}
}

\renewcommand{\shortauthors}{Drichel et al.}

\input{content/abstract.tex}




\maketitle

\input{content/introduction.tex}
\input{content/related_work.tex}
\input{content/evaluation_setup.tex}
\input{content/evaluation.tex}
\input{content/separating_dgas.tex}
\input{content/adaptive_retraining.tex}
\input{content/conclusion.tex}

\begin{acks}
This project has received funding from the European Union's Horizon 2020 research and innovation programme under grant agreement No 833418. 
Simulations were performed with computing resources granted by RWTH Aachen University under project rwth0438.
\end{acks}

\bibliographystyle{ACM-Reference-Format}
\bibliography{bibliography}

\appendix
\input{content/appendix.tex}

\end{document}

%% file: content/abstract.tex
\begin{abstract}

New malware emerges at a rapid pace and often incorporates Domain Generation Algorithms (DGAs) to avoid blocking the malware's connection to the command and control (C2) server.
Current state-of-the-art classifiers are able to separate benign from malicious domains (binary classification) and attribute them with high probability to the DGAs that generated them (multiclass classification).
While binary classifiers can label domains of yet unknown DGAs as malicious, multiclass classifiers can only assign domains to DGAs that are known at the time of training, limiting the ability to uncover new malware families.
In this work, we perform a comprehensive study on the detection of new DGAs, which includes an evaluation of 59,690 classifiers.
We examine four different approaches in 15 different configurations and propose a simple yet effective approach based on the combination of a softmax classifier and regular expressions (regexes) to detect multiple unknown DGAs with high probability.
At the same time, our approach retains state-of-the-art classification performance for known DGAs.
Our evaluation is based on a leave-one-group-out cross-validation with a total of 94 DGA families.
By using the maximum number of known DGAs, our evaluation scenario is particularly difficult and close to the real world.
All of the approaches examined are privacy-preserving, since they operate without context and exclusively on a single domain to be classified.
We round up our study with a thorough discussion of class-incremental learning strategies that can adapt an existing classifier to newly discovered classes.

\end{abstract}

%% file: content/introduction.tex
\section{Introduction}
\label{sec:introduction}

Distributed malware often needs to reconnect to the attacker to receive commands, exfiltrate data, or receive updates.
To this end, malware often incorporates Domain Generation Algorithms (DGAs) that make connection establishment robust against takedown attempts.
A DGA is a pseudo-random algorithm that generates a large set of domain names that the malware queries one by one to obtain the currently valid IP address of the command and control (C2) server.
The attacker knows the generation scheme of the DGA and can thus predict and register domains in advance.
Since the malware traverses all generated domains, most DNS queries result in non-existent domain (NXD) responses before the malware successfully resolves the C2 server's IP address.
Compared to using fixed domains or IP addresses, DGAs make it much more difficult to block communication between the attacker and the malware and make it easier for the attacker to switch C2 servers.

\copyrightnotice{\copyright\space Copyright held by the owner/author(s) 2022. This is the author's version of the work. It is posted here for your personal use. Not for redistribution. The definitive version was published in Proceedings of the 17th International Conference on Availability, Reliability and Security (ARES 2022), https://doi.org/10.1145/3538969.3538990} 

In the past, DGA binary detection classifiers have been proposed that are able to separate benign from DGA-generated domains with high probability and low false-positive rate (e.g.,~\cite{drichel_analyzing_2020,woodbridge_predicting_2016,yu_character_2018,schuppen_fanci_2018}).
Multiclass classifiers are even able to attribute malicious domains to the DGA that generated them, allowing for identification and targeted remediation of malware families (e.g.,~\cite{drichel_analyzing_2020,woodbridge_predicting_2016,tran2018lstm}).

While binary classifiers are partially able to label domains of yet unknown DGAs as malicious, multiclass classifiers can only attribute domains to DGA families that were known at training time and fail at detecting new, yet unknown DGAs.
New malware regularly appears in the real world, and new DGAs along with it.
A classifier that can identify yet unknown DGAs can accelerate the discovery of new malware and malware variants.

In this work, we conduct a comprehensive study on new DGA detection in which we examine the effectiveness of four different approaches.
In particular, we leverage different out-of-distribution (OOD) classification methods proposed in other application domains (e.g.,~\cite{cheon2019convolutional,abdelzad2019detecting,hendrycks17baseline}) that try to decide whether an input sample is in-distribution (ID), i.e., similar to the training data distribution, or OOD, i.e., different from the training data distribution, for new DGA detection.
In addition, we propose a method to detect new DGAs based on the combination of a softmax classifier and regular expressions (regexes).

For a successful new DGA classifier we set three requirements.

First, we require that a new DGA detection classifier retains the multiclass classification performance of a state-of-the-art classifier for known DGA classes.

Second, a new DGA classifier should not rely on training or tuning with OOD data since it is hard to define in advance and its selection can easily bias the learning of the classifier~\cite{hsu2020generalized}.

Third, we focus on approaches that operate without context, i.e., exclusively on the domain name to be classified, in order to protect the privacy of end users within an observed network as much as possible.
While eliminating potential privacy concerns, this paves the way for providing classification as a service.

We evaluate all classifiers in a five-fold cross-validated leave-one group-out evaluation using samples from all 94 DGA families available in DGArchive's open-source threat intelligence feed~\cite{plohmann_comprehensive_2016} to obtain meaningful results.
This results in an evaluation scenario that is particularly difficult and close to the real world, since the maximum number of known DGAs is used.
The more DGAs are included in the training data, the more difficult it is to correctly identify a DGA omitted from the training as a new DGA.
This is because it makes it more likely that the classifier will confuse the new DGA with a DGA present in the training data, as they could produce domains with similar properties.

In practice, samples of several unknown DGAs could be discovered and therefore samples may have to be divided into different classes.
To this end, we propose and evaluate an approach that clusters deep learning features extracted during classification and refines the clusters with additionally created context-less features.

After a new DGA is discovered, a new DGA detection classifier has to be able to adapt to the newly discovered class.
Therefore, we conclude our work with an analysis of the usefulness of incremental learning for the DGA detection use case.

%% file: content/related_work.tex
\section{Related Work}
\label{sec:related_work}

In this section, we first analyze approaches to new DGA detection proposed in the literature.
Subsequently, we present several new class detection methods from different application areas that could be used to detect new DGAs.

\subsection{Unknown DGA Detection}
The related work for unknown DGA detection can be split into two fields: (1) the correct classification of domains generated by unknown DGAs as malicious domains in a binary classification setting (e.g.,~\cite{schuppen_fanci_2018, drichel_analyzing_2020}), and (2) the detection of domains generated by unknown DGAs and their correct attribution to the unknown class in a multiclass classification setting (e.g.,~\cite{antonakakis_throwaway_2012,huang_detecting_2021}).

Dealing with unknown DGAs in the binary classification setting is easier as the main task is to separate benign from malicious domains. 
Malicious samples are not differentiated as to whether they were generated by an unknown DGA or by a known DGA.

For instance, Sch\"uppen et al.~\cite{schuppen_fanci_2018} proposed a system called FANCI which implements a random forest and a support vector machine classifier for DGA binary classification.
The authors evaluated FANCI's capability of detecting malicious samples generated by unknown DGAs in a leave-one-group-out cross-validation using 59 DGAs in total.
There, a classifier is trained using malicious domains from all DGAs but one, and its performance is evaluated on samples generated by the DGA that was left out.
On average, FANCI yielded a true positive rate (TPR) of approximately 96.389\% at a false positive rate (FPR) of 0.244\% and is thus highly capable of detecting unknown DGAs. 

In~\cite{drichel_analyzing_2020}, a similar evaluation was performed using different deep learning models and their performance was compared with FANCI.
On average, the deep learning models performed slightly better than FANCI and achieved TPRs between 96.961\% - 97.307\% at FPRs in the range of 0.134\% - 0.170\%.

In contrast, in the multiclass classification setting, domains are either attributed as benign or generated by a specific DGA.
Here, a distinction must be made in the case of malicious samples as to whether they were generated by an unknown DGA or by a known DGA.
Additionally, the classifiers may need to adapt to newly found classes.
In the context of DGA multiclass classification, two approaches to new DGA detection (\cite{antonakakis_throwaway_2012,huang_detecting_2021}) have already been proposed in the past, which we discuss in the following.

\subsubsection{Pleiades}
Antonakakis et al.~\cite{antonakakis_throwaway_2012} proposed Pleiades, a system capable of detecting machines infected with DGA-based malware within a monitored network. 
The system uses a combination of clustering and classification algorithms and relies heavily on DNS traffic tracking.
First, domain names as well as groups of machines that have queried these domains within a certain period of time are clustered based on 33 statistical features.
Alternating Decision Trees are then leveraged to map the generated clusters to known DGA families.
If a cluster cannot be assigned to any known DGA, the cluster is marked as a new DGA family.
Finally, a Hidden Markov Model is trained for each of the generated clusters and used to find active domains that are likely C2 domains of a particular DGA.

\subsubsection{Talos}
Huang et al.~\cite{huang_detecting_2021} proposed Talos, a context-less and deep learning based system for detecting new DGAs in the DGA multiclass classification setting.
Talos takes up the idea of leveraging feature vectors extracted from the early-layers of a deep neural network, which has proven to be good for detecting OOD samples in different application domains (e.g.,~\cite{lee2018simple,abdelzad2019detecting,cheon2019convolutional}), and performs simple clustering.

In detail, for each training sample, Talos extracts the feature vectors from the penultimate layer of a pre-trained neural network classifier during training.
Then, for each DGA family included in the training data, the feature vectors are averaged to calculate class cluster centers in the feature space.
The size of the clusters is determined by sorting the distances of all training samples of the respective classes by distance from their class centers and taking the 95th percentiles.
The final classifier is then represented by the pre-trained neural network, the class cluster centers, and the cluster sizes.
At inference time, the pre-trained neural network classifier is used to extract a feature vector for the sample to be classified.
If the feature vector falls within a cluster found during training, the class of the cluster that is closest to the feature vector is returned.
Otherwise, the sample is classified as belonging to a new DGA.

In practice, domains labeled as new DGA must be split into potentially multiple DGA families, since all samples that do not fall into any known class cluster are assigned to the same new DGA class.
However, the authors do not address this issue.

The used pre-trained neural network classifier is based on the Endgame multiclass classification model proposed in~\cite{woodbridge_predicting_2016} but incorporates an additional linear layer prior to the final classification layer.
From this added layer, Talos extracts the feature vectors used for classification.
The authors consider the additional layer as crucial for the detection of new DGAs.
However, this model modification leads to a significant drop in recall and F1-score for known DGA families compared to an unmodified one-vs-rest variant of the Endgame binary classification model~\cite{huang_detecting_2021}.

The authors evaluated Talos using a selection of ten DGA families, although the used data source~\cite{plohmann_comprehensive_2016} includes samples generated by over 90 DGAs.
In their evaluation, the authors concluded that Talos performs better with less known DGA families included in the training dataset, leaving unclear whether their approach works for more classes.
Therefore, it is questionable whether Talos is real-world applicable since more than 90 DGAs are already known.
An approach to new DGA detection is of practical interest only if it can handle all currently known DGAs while still detecting yet unknown DGA families.

\subsection{New Class Detection in other Domains}
Approaches to OOD classification have been proposed in several other domains in the past.

Hendrycks and Gimpel~\cite{hendrycks17baseline} proposed using probabilities from softmax distributions to separate ID from OOD samples since ID samples tend to have larger maximum softmax probabilities than OOD samples.
Liang et al.~\cite{liang2018enhancing} improve on Hendrycks and Gimpel's approach~\cite{hendrycks17baseline} by proposing the use of temperature scaling and adding small input perturbations which improve the separation of softmax score distributions between ID and OOD samples.	
Hsu et al.~\cite{hsu2020generalized} adapted the approach of Liang et al.~\cite{liang2018enhancing} by proposing to further decompose the confidence scores. Additionally, they removed the requirement to have OOD samples to fine-tune hyperparameters, thereby generalizing the approach.

Multiple works make use of feature vectors extracted from the early-layers of a deep neural network to separate ID from ODD samples (e.g.,~\cite{lee2018simple,abdelzad2019detecting,cheon2019convolutional}).
Lee et al.~\cite{lee2018simple} define a confidence score based on the Mahalanobis distance which is evaluated to conduct membership testing for given input samples.
Abdelzad et al.~\cite{abdelzad2019detecting} train one-class classifiers on the extracted features.
Cheon et al.~\cite{cheon2019convolutional} cluster the extracted features using the k-nearest-neighbors algorithm and evaluate the total squared distance between a feature vector and its k-nearest neighbors to decide whether a sample belongs to a known or an unknown class.

Other works (e.g.,~\cite{fan2004using,sabokrou2018adversarially,lee2018training}) generate artificial OOD samples that are used to improve understanding of ID sample distribution and constrain its boundary, thereby improving OOD sample detection. In~\cite{sabokrou2018adversarially,lee2018training}, Generative Adversarial Networks (GANs) are leveraged for this purpose.

%% file: content/evaluation_setup.tex
\section{Evaluation Setup}
\label{sec:evaluation_setup}
In this section, we present our evaluation setup including the data sources used, the selected state-of-the-art DGA classifier, the selected new DGA detection approaches being evaluated, and the methodology.

\subsection{Data}
For our evaluation, we obtain malicious and benign labeled data from two different data sources.

\paragraph{Malicious Data}
Our source for malicious data is DGArchive~\cite{plohmann_comprehensive_2016}, which contains domains generated by reimplementations of known DGAs and seeds.
For our evaluation, we use all available samples up to 2020-09-01. 
In total, we obtained approximately 126 million domains generated by 94 different DGA families.

\paragraph{Benign Data}
\label{sec:benign_data}
We obtain benign labeled data from the central DNS resolver of the campus network of RWTH Aachen University which includes several academic as well as administrative networks, networks from student residences, the network of the affiliated university hospital, and eduroam~\cite{eduroam}.
In order to remove potential DGA generated domain names from our benign data, we filter all recorded domains against all DGA samples known in DGArchive.
In total, we obtained approximately 26 million unique NXDs for DNS queries during the whole month of September 2019.

\begin{figure*}
	\centering
	\includegraphics[width=0.95\linewidth]{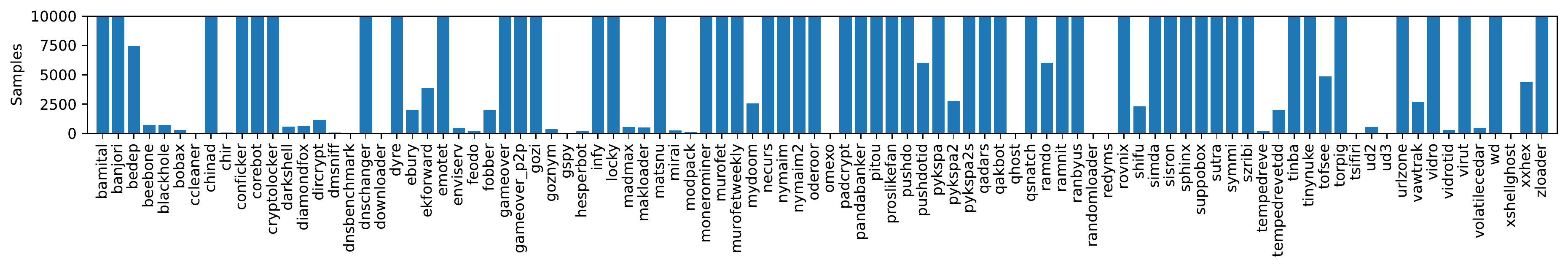}
	\caption{Dataset sample distribution per included DGA family.}
	\label{fig:sample_distribution}
\end{figure*}

\paragraph{Dataset}
\label{ref:dataset}
We create a dataset for our evaluation by random sub-sampling 10,000 samples for each DGA class included in DGArchive and the benign class.
We include all available samples for DGA classes for which fewer than 10,000 samples are known.
This results in a dataset containing 557,764 samples generated by 94 different DGA families and additional 10,000 random benign samples obtained from the real-world network.
We present the sample distribution for each DGA family included in the dataset in Fig~\ref{fig:sample_distribution}.

\subsection{Selected State-of-the-Art Classifier}
Our goal is to propose an approach to new DGA detection while retaining or even improving the performance of state-of-the-art DGA multiclass classifiers.
We therefore use the currently best DGA multiclass classifier as a basis.
The authors of~\cite{drichel_analyzing_2020} propose a residual neural network (ResNet) based classifier for DGA multiclass classification.
In a comparative evaluation including convolutional (CNNs)~\cite{yu_character_2018} and recurrent neural networks (RNNs)~\cite{woodbridge_predicting_2016,tran2018lstm}, the authors show that their ResNet-based classifier outperforms the other approaches and is able to correctly attribute samples to multiple DGA families, which the other classifiers cannot correctly attribute.

The ResNet classifier consists of an embedding layer, eleven residual blocks, and lastly a dense layer.
The embedding layer incorporates semantics into the character-wise encoding of an input domain name.
The residual blocks contain two stacked one-dimensional convolution layers whose output is combined with the unmodified input to the block. 
These so-called skip connections facilitate training of the neural network as they allow the gradient to bypass certain layers unaltered and counteract the vanishing gradient problem.
The output of each residual is then processed by applying the ReLU activation function and then performing a max pooling operation.
The final dense layer consists of as many nodes as there are classes in the training data and uses the softmax activation function to perform the final multinomial logistic regression.
We use the same domain name pre-processing to encode domains as proposed by the model authors.
More detailed information about the classifier used can be found in~\cite{drichel_analyzing_2020}.

\subsubsection{Deep Learning Model as Feature Extractor}
\label{ref:dl_feature_extraction}
We use the early-layer output of the ResNet classifier to extract features since several approaches to OOD detection rely on them.
After training the ResNet classifier, the training samples are again fed into the model and the intermediate outputs are collected at multiple locations in the network.
We start by capturing the output of the embedding layer and end with the dense layer output but before the softmax activation.
In between, we capture the output of each residual block before ReLU activation, after ReLU activation, and after the max pooling operation.
In total, we collect the intermediate outputs at 31 network locations.

\subsection{Selected Approaches to New DGA Detection}
In this work, a new DGA classifier is a classifier that can classify an input domain either as benign, to any known DGA class, or as a new DGA.
Any sample that is neither benign nor associated with any of the known DGAs should receive the label \textit{new DGA}.
This gives samples from multiple unknown DGAs the same \textit{new DGA} label and hence it may be necessary to subdivide domains into multiple classes.

Here we focus on context-less classifiers to preserve privacy as much as possible and reduce potential concerns about delivering classification as a service.
Thus, we do not further investigate Pleiades~\cite{antonakakis_throwaway_2012} as it relies on extensive tracking of DNS traffic.

In addition, we focus on approaches that do not require access to OOD samples.
Hence, we do not include the temperature scaling proposed by Liang et al.~\cite{liang2018enhancing} or the approach based on the Mahalanobis distance by Lee et al.~\cite{lee2018simple} because both approaches make use of OOD validation sets to tune hyperparameters.

Furthermore, since we strive to retain or even improve the performance of the state-of-the-art DGA multiclass classifier used, we exclude the input pre-processing methods proposed in~\cite{liang2018enhancing,hsu2020generalized}.
The problem is that these input pre-processing methods require calculating the gradient for each sample up to the input.
However, the used deep learning model uses a non-differentiable embedding layer, thus the proposed input pre-processing methods are not directly applicable.

In the following, we present the baseline and the different approaches investigated to detect new DGAs.

\subsubsection{Baseline}
\label{sec:baseline}
All approaches to new DGA detection are compared against the \textit{Baseline} to evaluate their performance.
The \textit{Baseline} is a simple multiclass deep learning DGA classification model with no new DGA detection capabilities.
If a DGA sample from an unknown class is presented to the classifier, it will inevitably be misclassified as either benign or a known DGA.
Thus this classifier is just a conventionally trained deep learning model without any additional logic.

\subsubsection{Regex-based Approach}
A simple approach to detect and possibly correct errors in predictions of a trained deep learning model can be implemented using regular expressions (regexes).
When a sample is assigned to a known DGA family, a regex can be used to check correctness, correct classification errors, or categorize the sample as belonging to a new DGA.
To this end, we generate simple regexes for all known DGA families that match as closely as possible all available samples from each class.
These regexes check sample length, alphabet set used, presence of prefixes and suffixes, including top-level domains (TLDs) and public suffixes used.
For instance, the regular expression for the \textit{fobber} DGA is \mbox{[a-z]\{10,17\}\textbackslash.(com$\vert$net)}, i.e., \textit{fobber} generates strings of length 10 to 17, using only the characters from a to z, and then appends the period and one of the two possible TLDs (com or net), giving the final domain name.
Based on regexes, we elaborate two regex-based classifiers.

\paragraph{Regex - Error Detection Classifier}
The first approach performs error detection: if a domain name is attributed to a known DGA family, the regex extracted for that DGA class is used to check if the sample can be generated by that particular DGA.
If the regex matches the class returned by the deep learning model, it is accepted, otherwise the prediction is rejected and the sample is labeled as belonging to a new DGA class.
When the deep learning classifier outputs the benign label, the prediction is always accepted.

\paragraph{Regex - Error Correction Classifier}
The second approach tries to correct prediction errors using regexes.
For this purpose, the top $k$ scoring labels from the softmax vector returned by the deep learning classifier are checked using regexes.
The top $k$  classes are matched against the corresponding class regexes in descending order.
The first class matching the regex is returned.
If there is no regex match in the top $k$  classes, the sample is labeled as new DGA.
The benign class is treated the same as the malicious DGA classes, using a match-all regex.
We evaluate this approach using up to five classes to be tested ($k\in {2,3,4,5}$).
$k=1$ corresponds to the \textit{Regex - Error Detection} approach.

\subsubsection{Max-Softmax}
The \textit{Max-Softmax} approach is based on the idea of~\cite{hendrycks17baseline}.
For training, a deep learning model is trained conventionally and then the entire training set is run through the trained classifier again.
For each class in the training set, an individual threshold is calculated using the softmax scores of the correctly predicted training samples of each class.
During inference, a sample is fed into the deep learning classifier and the softmax confidence score corresponding to the predicted label is compared to the predicted class threshold.
If the softmax confidence score is lower than the threshold for the class, the sample is flagged as a new DGA.
If the threshold is exceeded, the prediction is the label returned by the deep learning model.
We evaluate this approach using the 5th and 10th percentiles of the softmax score distribution of the correctly classified samples for each class as thresholds.
Preliminary testing indicated that smaller thresholds yield better results.
Thus, we also evaluate the minimum softmax score per class $Class\textsubscript{Min}$, $0.95 \cdot Class\textsubscript{Min}$, and $0.90 \cdot Class\textsubscript{Min}$ as class thresholds.

\subsubsection{Family Vector}
\label{sec:fv}
This approach leverages the idea of clustering the feature vectors extracted from the early-layers of a deep neural network presented in~\cite{huang_detecting_2021}.
For training, a conventionally trained deep learning classifier is fed the entire training set one more time and the output from the early-layers is collected (cf. Section~\ref{ref:dl_feature_extraction}).
A cluster for each layer and class in the training set is learned by averaging the feature vectors of each class.
In addition, a class distance threshold is calculated for each cluster, representing the size of the cluster.
In this work, we evaluate two possible approaches for threshold setting, \textit{Mean Threshold} and \textit{Median Threshold}.
The \textit{Mean Threshold} corresponds to the 95th percentile of the class distance distribution, while the \textit{Median Threshold} is based on the idea of the median.
When predicting a label for a domain, the layer output is extracted using the deep learning model.
If the resulting feature vector falls within one of the learned clusters, the label of the closest cluster that does not exceed the threshold is returned.
If the domain's feature vector lies outside of all clusters, the sample is flagged as belonging to a new DGA class.
For each of the 31 network locations of the ResNet model from which we collect the early-layer output, we train an individual classifier.

\begin{figure}
	\centering
	\includegraphics[width=0.95\linewidth]{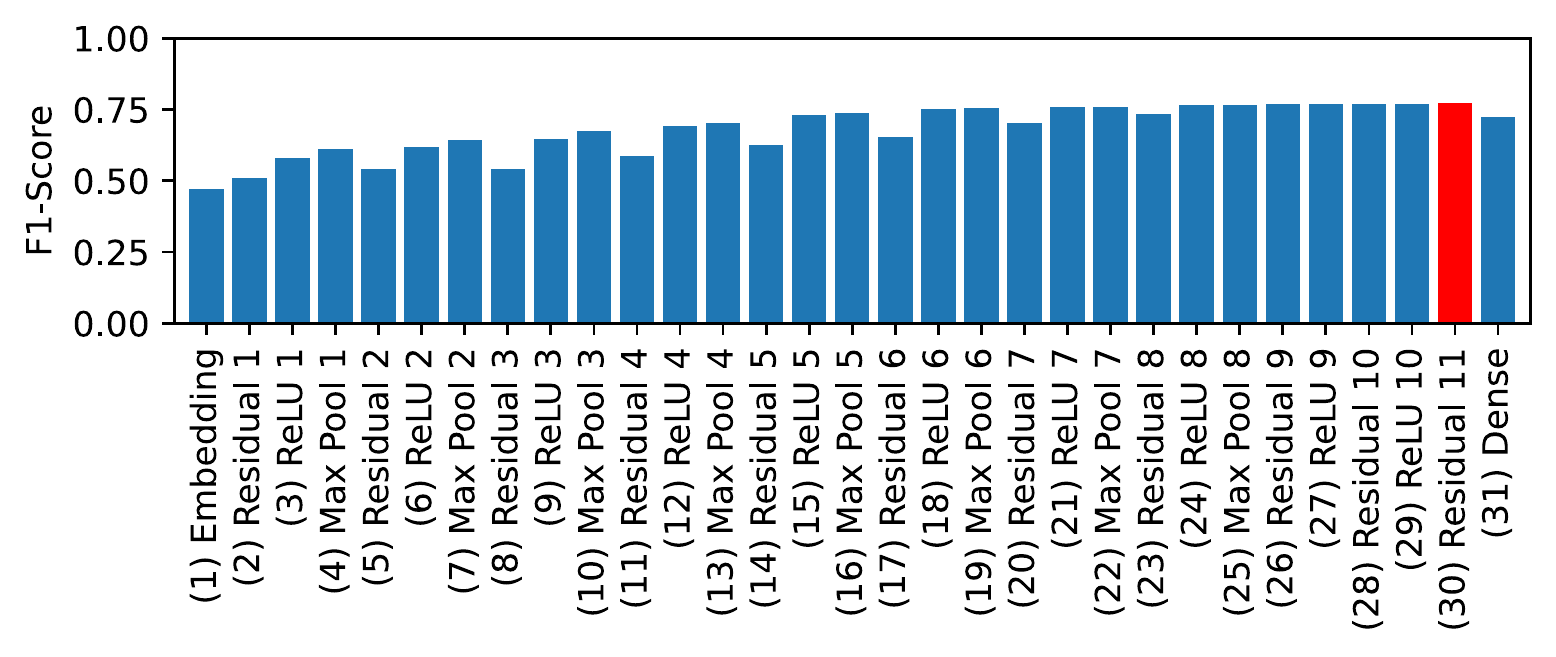}
	\caption{F1-scores obtained by the Family Vector approach with mean thresholding per layer classifier.}
	\label{fig:layers}
\end{figure}

In Fig.~\ref{fig:layers}, we anticipate the evaluation results slightly in order to visualize the individual layer classifiers.
For each trained family vector classifier with mean thresholding, we display the achieved F1-score, which is used as a metric to assess a classifier's overall performance.
The best layer classifier is colored red.

\subsubsection{One-Class Classifier}
This approach is based on~\cite{abdelzad2019detecting}, which uses a one-class classifier for OOD detection, using the early-layer outputs of a sample to be classified as features.
Similar to the \textit{Family Vector} approach, a conventionally trained deep learning classifier is used to generate feature vectors from the training data for each neural network layer.
The outputs of each early layer are used as feature vectors to train an one-class classifier in a second step.
During inference, the sample class is classified using the deep learning model.
Then the feature vector is extracted and fed to the corresponding one-class classifier.
If the one-class classifier classifies the sample as ID, the label of the deep learning model is returned, otherwise the sample is labeled as belonging to a new DGA class.
In this work, we evaluate a \textit{one-class support vector machine}, an \textit{isolation forest}, and a \textit{local outlier factor} classifier.

The feature vectors extracted from the first layers are high dimensional, which posed a problem for our parallelized evaluation.
In order to cover the memory needs of the one-class classifiers with the 180 GB available memory of our test machine, we had to reduce the maximum number of training samples per class to 1,000 and exclude the first layers up to the output of the fifth residual of the ResNet model.

\paragraph{Naive Layer Ensembles}
The \textit{One-Class Classifier} and the \textit{Family Vector} approach train an individual classifier for each network location of the ResNet model from which we collect early-layer features.
It is thus possible to combine the individual layer classifiers into layer ensemble classifiers in order to improve the classification performance by combining the information obtained from different layers.
For the \textit{One-Class Classifier} ensemble we combine all evaluated layer classifiers into an ensemble classifier. 
For the \textit{Family Vector} ensemble, we perform a simple layer section in which we combine all the layers before and after the best layer until the layer's F1-score is 0.05 lower than the maximum score.
Thereby, the \textit{Family Vector} ensemble includes eleven layer classifiers.
We use majority voting to combine the individual classification results to a final prediction.
In case of a tie, the sum of the prediction confidence scores is used as a tie-breaker.

\subsubsection{Discarded Approach: k-Nearest-Neighbors Classifier}
This approach is based on the k-Nearest-Neighbors (kNN) algorithm and proposed in~\cite{cheon2019convolutional}.
Again, feature vectors are extracted using a conventionally trained deep learning model and its training data.
However, only feature vectors of the training samples correctly classified by the trained network are used for training here.
For each class, the sigmoid activation function is applied to each feature vector and the total squared distance of each sample to the $k$ nearest neighbors in the same class are calculated.
For each cluster, the 90th percentile of the total squared distance distribution is used as the cluster threshold.
During inference, the sample's feature vector is extracted and then the total squared distance to all clusters is calculated.
If the distance to a cluster is smaller than the threshold of the corresponding class, the sample is classified as ID and the prediction of the deep learning model is returned.
If all cluster thresholds are exceeded, the sample is labeled as  new DGA.

Preliminary tests have shown that the kNN-based classifier is significantly slower compared to the \textit{Family Vector} approach.
The main reason for this is the many distance calculations required during training and inference.
During training, the distances of each feature vector to all other feature vectors of the same class must be calculated.
During inference, the distances between the sample's feature vector and the feature vectors of all training samples must be calculated.
This is a multiple of the distance calculations required for the \textit{Family Vector} approach.
In addition, this approach achieved significantly poorer results in preliminary tests compared to other approaches.
Based on these facts, we discard this approach. 

\subsection{Methodology}
Below we present our evaluation scenario and the evaluation metrics used to assess the different approaches.

\subsubsection{Evaluation Scenario}
In order to obtain meaningful results, we conduct a five-fold cross-validated leave-one-group-out evaluation for each approach examined.
To this end, the complete dataset (cf. Section~\ref{ref:dataset}) is split into five folds stratified over all included classes.
In each fold, each DGA class is left out once from the training data, i.e., the samples of the omitted DGA class are removed from the training set while remaining in the testing set.
If there are fewer than the maximum number of 2,000 testing samples for a left out DGA, we populate the test set with randomly chosen samples that were previously excluded from the training data of a fold.
If less than 2,000 samples are known for a DGA class, we use all available samples.
Note that the training and testing sets are fully disjoint, thus this does not result in using training samples for testing.
Since each DGA of the 94 DGA classes is left out once and the evaluation is performed five-folded, we perform 94$\cdot$5 = 470 evaluations per new DGA detection classifier and parameterization.
In total, we evaluate 59,690 new DGA detection classifiers and perform an additional 2,350 ensemble evaluations.
The conventionally trained deep learning model used by all new DGA detection approaches is always trained using early stopping with a patience of three epochs to avoid overfitting. During training, the model's performance is assessed using random 5\% splits of the training data used.

This evaluation scenario allows us to (1) measure classification performance in detecting new DGAs (DGAs that were not present in the training data), and (2) ensure that the added new DGA detection capability does not negatively affect the detection performance of known DGAs (as we also evaluate a classifier using test samples of known DGAs).

\subsubsection{Evaluation Metrics}
\label{sec:evaluation_metrics}
In order to assess the different approaches to new DGA detection, we use the precision, recall, and F1-score.
The precision measures the fraction of true positives that are labeled as positive by a classifier, while the recall measures the fraction of correctly retrieved positives and equals the true positive rate. The F1-score is the harmonic mean of precision and recall. 
We use macro-averaging to obtain the final evaluation scores, i.e., we average the scores for the individual classes without considering the actual number of samples available per class.
Thereby, we give each DGA class the same level of importance independent of the number of samples per class inside the test set. 
Thus, the final scores are not biased towards DGA classes that are well-represented in the test set.

In order to be able to assess the ability to detect new and known DGAs independently, we separate the evaluation metrics.
The precision, recall, and F1-score without any prefix are used to measure the detection performance of known DGAs. For these scores, the individual class scores for all DGAs included in the training set are averaged. 
The Left-Out-Class (LOC) precision, LOC recall, and LOC F1-score metrics are calculated by averaging the individual class scores for the DGAs that were left out of training. Thus, these scores measure the detection capability of new DGAs.

%% file: content/evaluation.tex
\section{Evaluation}
\label{sec:evaluation}
In this section, we present our evaluation results.
For the One-Class Classifier and Family Vector approaches for which multiple layer classifiers are trained, we present the results for the best layer classifier as well as for the ensemble classifiers.

\subsection{Results}
In Table~\ref{tab:results}, we show for each individual approach and parameterization examined the averaged F1-score, precision, and recall values for the known classes (i.e., for the DGA classes for which samples were included during training of the classifier) and the averaged LOC F1-score, LOC precision, and LOC recall values for the unknown classes (i.e., for DGAs for which no samples were included during training) (cf. Section~\ref{sec:evaluation_metrics}).

\input{tables/results.tex}

The results of the Baseline (cf. Section~\ref{sec:baseline}) are displayed at the top of Table~\ref{tab:results}.
As the Baseline has no new DGA detection capabilities, the samples of the left out classes are bound to be misclassified, so the LOC scores are all zero.

Only the regex-based and the Max-Softmax approaches up to a threshold of $Class\textsubscript{Min}$ (the minimum softmax score per class observed during training) achieve better F1-scores than the Baseline.
Therefore, only these approaches are of particular interest as they are able to retain or even improve the performance of state-of-the-art DGA classification for known classes.

Regex - Error Detection achieves a LOC F1-score of 31.829\%, which is the best of all approaches.
Regex - Top $k$ Error Correction approaches worsen in LOC F1-score and improve in F1-score (except $k=5$, very close to $k=4$) with  increasing $k$.
This can be explained by the classifier getting multiple attempts to correct its prediction.
However, this increases the likelihood that an unknown DGA sample will be mistakenly assigned to a known class, since the regexes used are simple and have a high degree of overlap.
Regex - Top 4 Error Correction achieves the best F1-score among all approaches.
Comparing its LOC F1-score with the one of Regex - Error Detection indicates that the small increase in F1-score might not outweigh the loss of 10.897\% in LOC F1-score.

The Max-Softmax approaches with a threshold up to $Class\textsubscript{Min}$ yield better F1-scores than the Baseline, but no better F1-scores or LOC F1-scores than the regex-based approaches.

All other approaches achieve F1-scores worse than the Baseline.
The best One-Class Classifier approach is based on the local outlier factor classifier while Mean Threshold outperforms Median Threshold in the Family Vector approach.

The ensemble One-Class Classifier approaches based on the isolation forest and the support vector machine achieve better F1-scores than the best single-layer classifiers (still worse than the Baseline), but worse LOC F1-scores.
The ensemble based on the local outlier factor gives results similar to the best single-layer classifier.
The Family Vector ensembles achieve worse results than their single-layer counterparts.
Since the ensemble results are usually worse or, in the case of the local outlier factor classifiers, achieve similar LOC F1-scores but require a multiple of the computing power, we discard all ensemble classifiers in the following.

In summary, the best approach for new DGA detection that retains state-of-the-art detection performance for known DGA classes is Regex - Error Detection, which yields a LOC F1-score of 31.829\%.

\subsection{Analysis per Omitted DGA Family}
\begin{figure*}
	\centering
	\begin{subfigure}{0.95\linewidth}
		\centering
		\includegraphics[width=\linewidth]{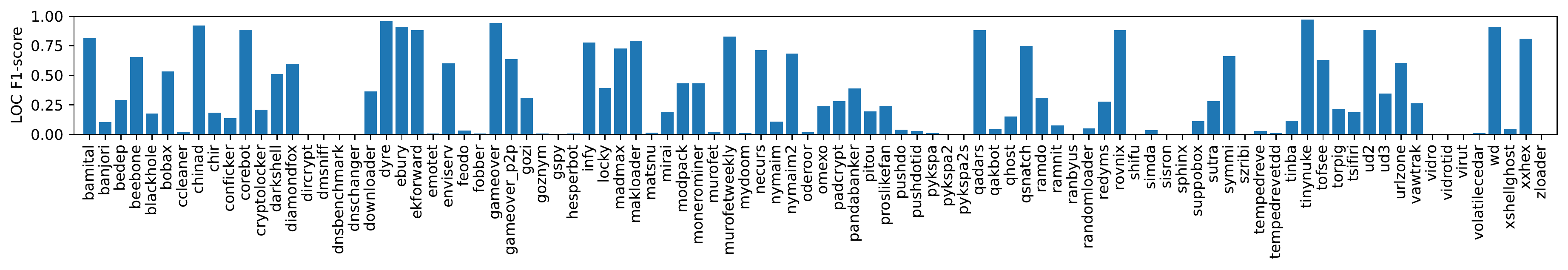}
		\caption{Regex - Error Detection}
		\label{fig:regex_error_detection}
	\end{subfigure}
	\begin{subfigure}{0.95\linewidth}
		\centering
		\includegraphics[width=\linewidth]{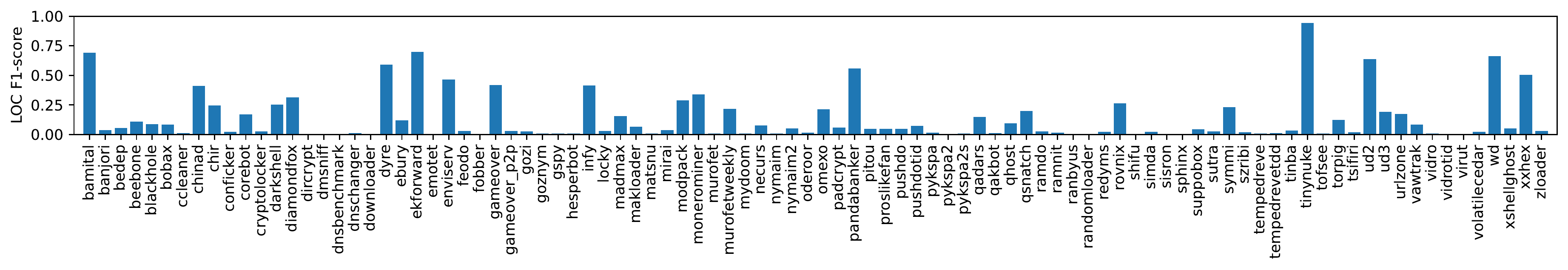}
		\caption{Max-Softmax - Threshold: $Class\textsubscript{min}$}
		\label{fig:max_softmax}
	\end{subfigure}
	\begin{subfigure}{0.95\linewidth}
		\centering
		\includegraphics[width=\linewidth]{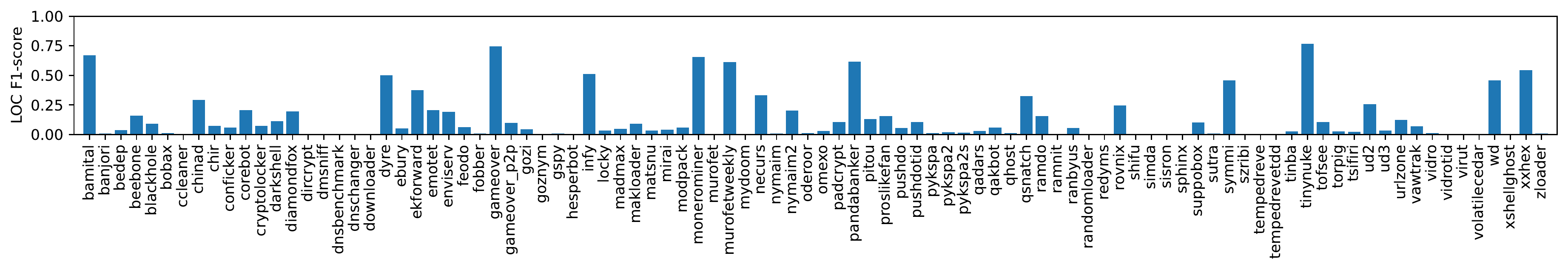}
		\caption{Family Vector - Mean Threshold}
		\label{fig:family_vector_mean}
	\end{subfigure}
	\begin{subfigure}{0.95\linewidth}
		\centering
		\includegraphics[width=\linewidth]{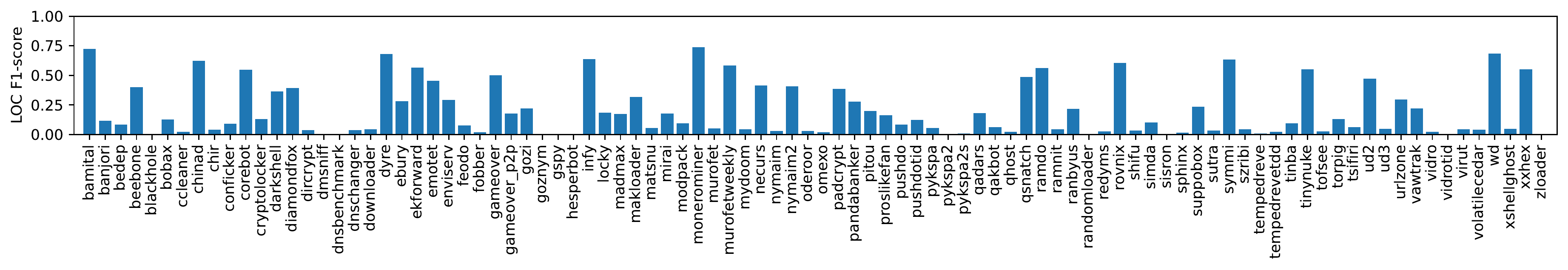}
		\caption{One-Class Classifier - Local Outlier Factor}
		\label{fig:occ_lof}
	\end{subfigure}
	\caption{LOC F1-score distributions of best performing new DGA detection classifiers.}
	\label{fig:loc_graphs}
\end{figure*}

In the following, for the best configuration of each approach, we analyze the ability to detect new DGAs individually for each DGA family excluded from training.
In Fig.~\ref{fig:loc_graphs}, we visualize the five-folded LOC F1-scores for each DGA family and for the best configuration of each approach that achieves the highest LOC F1-score.

The Regex - Error Detection approach is the best and is able to correctly label samples of multiple DGAs, that have been excluded from training, as new DGAs with a high probability.
This approach is able to correctly attribute 14 out of 94 DGA classes with F1-scores above 80\%, 29 DGAs above 50\%, and 46 DGAs above 20\%.

Compared to the other three approaches, DGAs that are well recognized by Max-Softmax, Family Vector, or the One-Class Classifier approach are mostly also well recognized by Regex - Error Detection.
The biggest difference can be seen in the detection of five DGAs: \textit{emotet}, \textit{monerominer}, \textit{pandabanker}, \textit{ramdo}, and  \textit{ranbyus}.
These five DGAs are better detected in at least one approach compared to Regex - Error Detection.
This can be explained by the fact that these DGAs generate similar domains as DGAs that were included in the training of the deep learning model and that the regexes used overlap significantly.
For instance, the \textit{emotet} DGA is mostly confused (>90\% of all samples) with the \textit{ramnit} DGA.
The regex for the \textit{emotet} DGA is \mbox{[a-y]\{16,16\}\textbackslash.(eu)} and the regex for the \textit{ramnit} DGA is \mbox{[a-y]\{16,16\}\textbackslash.(bid$\vert$click$\vert$com$\vert$eu)} and thus matches 100\% of all \textit{emotet} samples.
Since the \textit{ramnit} DGA class was included during the training of the deep learning model, and both DGAs generate similar domains, it is not surprising that the deep learning model confuses samples of these two classes. 
In combination with the overlapping regexes, the identification of the new DGA class fails.
In comparison, the Family Vector and the One-Class Classifier approach are partially able to attribute \textit{emotet} samples correctly.
We reckon that the early-layer outputs leveraged by these two approaches contain features that partly allow to separate the generation process of these two classes.

The Max-Softmax approach, on the other hand, is only significantly better at detecting the \textit{pandabanker} DGA and worse at detecting most other DGAs compared to Regex - Error Detection.

Other DGAs that are poorly recognized by Regex - Error Detection are also poorly detected in the other approaches.
Note, missing bars within the figures do not mean that samples of those classes are attributed to the benign class but rather to a different DGA.
The averaged F1-score of the benign class is 98.289\% for the Regex - Error Detection approach.
For instance, related DGA families such as \textit{pykspa}, \textit{pykspa2}, and \textit{pykspa2s} generate similar domains.
Therefore, the deep learning model extracts similar feature values in early layers and the generated regexes are identical or at least heavily overlapping.~\footnote{The regexes of \textit{pykspa2} and \textit{pykspa2s} are identical. The regex of \textit{pykspa} has a shorter maximal domain length of 12 (instead of 15) and does not include the cc and biz TLD.}
Hence, it is not surprising that related DGA classes are confused.

Furthermore, multiple DGAs even generate identical domains, making contextless separation practically impossible.
For instance, in the open-source threat intelligence feed of DGArchive~\cite{plohmann_comprehensive_2016}, 79.092\% of the unique domains generated by the \textit{oderoor} DGA are also generated by the \textit{vidro} DGA~\cite{drichel_making_2020}.

\input{tables/statistics.tex}

In Table~\ref{tab:statistics}, we display statistical data over the LOC F1-score distributions.
It shows that the Regex - Error Detection approach achieves the highest values for the average, median, and maximum LOC F1-scores.
All approaches fail to detect a few DGAs for the previously mentioned reasons, hence the minimum LOC F1-score for all approaches is zero.
The standard deviation is highest for the regex-based approach because it recognizes the most classes with higher probabilities.
The Regex - Error Detection approach achieves a median of 19.444\%.
This roughly corresponds to the approach being able to correctly label every fifth domain out of half of the analyzed DGA families as belonging to a new DGA.
At the same time, it retains state-of-the-art classification performance for known DGAs.

In a real-world application, the identification of a new DGA would not be based on the classification of a single query, but rather on multiple classifications.
Since DGAs tend to generate a large amount of domain names, even an F1-score of 19.444\% is helpful in identifying new DGAs.
In this work, the new DGA classifiers operate context-less in order to preserve privacy as much as possible.
However, a third party that uses this classification as a service could also exploit context information.
Correlating queried domains labeled as belonging to new DGAs with the hosts that queried those domains could greatly improve the correct detection of new DGAs.

%% file: tables/results.tex

\begin{table*}[!t]
	\caption{Averaged results of five-folded leave-one-group-out evaluation.}
	\label{tab:results}
	\centering
		\begin{tabular}{ll|ccc|ccc}
			\hline \Tstrut
			\multirow{2}{*}{\textbf{Approach}} & \multirow{2}{*}{\textbf{Configuration}} & \multirow{2}{*}{\textbf{F1-Score}} & \multirow{2}{*}{\textbf{Precision}} & \multirow{2}{*}{\textbf{Recall}} & \textbf{LOC} & \textbf{LOC} & \textbf{LOC} \\
			& & & & & \textbf{F1-Score} & \textbf{Precision} & \textbf{Recall} \\
			\hline \Tstrut
			Baseline & - & 0.79237 & 0.80677 & 0.80291 & 0.00000 & 0.00000 & 0.00000 \\
			\hline \Tstrut
			\multirow{5}{*}{Regex} & Error Detection & 0.80110	& 0.82007 & 0.80634 & \textbf{0.31829} & 0.42606 & 0.32568 \\
			& Top 2 Error Correction & 0.80229	& 0.82096 & 0.80844 & 0.28883 & 0.52792 & 0.24804 \\
			& Top 3 Error Correction & 0.80279	& 0.82161 & 0.80918 & 0.24783 & \textbf{0.52994} & 0.20132 \\
			& Top 4 Error Correction & \textbf{0.80320}	& 0.82210 & 0.80976 & 0.20932 & 0.51229 & 0.16384 \\
			& Top 5 Error Correction & 0.80315	& 0.82209 & \textbf{0.80983} & 0.18075 & 0.49585 & 0.13673 \\
			\hline \Tstrut
			\multirow{5}{*}{Max-Softmax} & Threshold: $0.90 \cdot Class\textsubscript{Min}$ & 0.79471 & 0.81472 & 0.80115 & 0.09803 & 0.29442 & 0.07139 \\
			& Threshold: $0.95 \cdot Class\textsubscript{Min}$ & 0.79475 & 0.81538 & 0.80041 & 0.10842 & 0.28661 & 0.08225 \\
			& Threshold: $Class\textsubscript{Min}$ & 0.79393 & 0.81655 & 0.79754 & 0.13228 & 0.27480 & 0.11463 \\
			& Threshold: 5th percentile & 0.77867 & 0.83075 & 0.75772 & 0.10580 & 0.06350 & 0.43236 \\
			& Threshold: 10th percentile & 0.75922 & \textbf{0.83847} & 0.71914 & 0.08269 & 0.04632 & \textbf{0.52104} \\
			\hline \Tstrut
			\multirow{2}{*}{Family Vector} & Mean Threshold & 0.77188 & 0.78646 & 0.79351 & 0.12927 & 0.12889 & 0.16589 \\
			& Median Threshold & 0.76016 & 0.79603 & 0.76068 & 0.09250 & 0.05717 & 0.32790 \\
			\hline \Tstrut
			\multirow{3}{*}{One-Class Classifier} & Isolation Forest & 0.74335 & 0.79209 & 0.74394 & 0.01443 & 0.00876 & 0.05794 \\
			& Support Vector Machine & 0.72998 & 0.79987 & 0.70332 & 0.04438 & 0.02625 & 0.21126 \\
  			& Local Outlier Factor & 0.78886 & 0.81353 & 0.79062 & 0.20195 & 0.16501 & 0.35987 \\
			\hline \Tstrut
			\multirow{1}{*}{Ensemble} & Mean Threshold & 0.76782 & 0.78149 & 0.79144 & 0.11962 & 0.11699 & 0.15205 \\
			\multirow{1}{*}{Family Vector} & Median Threshold & 0.75827 & 0.79306 & 0.76033 & 0.09199 & 0.05682 & 0.32233 \\
			\hline \Tstrut
			\multirow{2}{*}{Ensemble} & Isolation Forest & 0.78790 & 0.80638 & 0.79583 & 0.00015 & 0.00031 & 0.00010 \\
			\multirow{2}{*}{One-Class Classifier}& Support Vector Machine & 0.76133 & 0.80734 & 0.74698 & 0.03111 & 0.01936 & 0.10454 \\
			& Local Outlier Factor & 0.78811 & 0.81386 & 0.78896 & 0.20344 & 0.16192 & 0.37815 \\
			\hline
		\end{tabular}
\end{table*}

%% file: tables/statistics.tex

\begin{table}[!t]
	\caption{Statistics of the LOC F1-score distributions for the best configurations of each approach.}
	\label{tab:statistics}
	\centering
	\resizebox{\columnwidth}{!}{
	\begin{tabular}{lccccc}
		\toprule
		\multirow{2}{*}{\textbf{Approach}} & \multicolumn{5}{c}{\textbf{LOC F1-Score}}  \\ \cline{2-6} \Tstrut
		& \textbf{Average} & \textbf{Median} & \textbf{Min} & \textbf{Max} & \textbf{Stddev} \\
		\midrule
		Regex - & \multirow{2}{*}{0.31829} & \multirow{2}{*}{0.19444} & \multirow{2}{*}{0.00000} & \multirow{2}{*}{0.97082} & \multirow{2}{*}{0.33208} \\ 
		Error Detection & & & & & \\
		\midrule
		Max-Softmax - & \multirow{2}{*}{0.13228} & \multirow{2}{*}{0.03446} & \multirow{2}{*}{0.00000} & \multirow{2}{*}{0.94369} & \multirow{2}{*}{0.19841} \\ 
		Threshold: $Class\textsubscript{Min}$ & & & & & \\
		\midrule
		Family Vector - & \multirow{2}{*}{0.12927} & \multirow{2}{*}{0.04303} & \multirow{2}{*}{0.00000} & \multirow{2}{*}{0.76819} & \multirow{2}{*}{0.19264} \\
		Mean Threshold & & & & & \\
		\midrule
		One-Class Classifier - & \multirow{2}{*}{0.20195} & \multirow{2}{*}{0.09423} & \multirow{2}{*}{0.00000} & \multirow{2}{*}{0.73717} & \multirow{2}{*}{0.22023} \\
		Local Outlier Factor& & & & & \\
		\bottomrule
	\end{tabular}
	}
\end{table}

%% file: content/separating_dgas.tex
\section{Separating Multiple DGAs}
\label{sec:separating_multiple_dgas}

When using any of the proposed new DGA classifiers in the real world, samples of several unknown DGAs could be discovered.
Therefore, it may be necessary to split samples identified as belonging to new DGAs into multiple classes.
To this end, we propose to cluster the early-layer features extracted for domains labeled as new DGA and subsequently refine the found clusters by leveraging additional context-less features.
We evaluate this approach by classifying all 26 million unique and unfiltered samples obtained from our benign data source (cf. Section~\ref{sec:benign_data}), demonstrating the feasibility of new DGA detection using real-world data.

For this purpose, we create a similar training dataset as described in Section~\ref{sec:benign_data}, but revert the malicious data to the status of 2019-09-01 in order to generate a realistic training dataset since the benign labeled samples are from September 2019.
As a result, we lose all samples from three DGAs that were not known at the time.
We then train a Regex - Error Detection approach and classify all unfiltered samples from our benign data source, resulting in 110,284 unique samples (0.423\%) being labeled as belonging to new DGAs.

In order to separate the new DGA samples from obvious false positives that are generated by misconfigured/outdated software or by intentional misuse of the DNS, e.g., by anti-virus software~\cite{drichel_making_2020}, we remove all samples that do not have a unique second-level domain and public suffix (TLD) combination.
This does not result in samples of new DGAs being removed as they typically generate unique second-level domains to avoid simple blocklisting.
This process reduces the total number of samples to 63,646 (0.244\%).

For these samples, we extract the early-layer features from the penultimate layer of the trained ResNet model.
We chose features extracted from this layer because the Family Vector approach worked best with them, and because this layer is further down the network, it produces less dimensional features that facilitates subsequent clustering.
We use the X-Means~\cite{pelleg2000x} algorithm for clustering the extracted features which extends the K-Means algorithm by an efficient estimation of the number of clusters.
However, the estimated number of clusters is not optimal because, e.g., samples that actually belong to the same cluster can be separated due to different domain lengths.
Therefore, manual inspection of clusters by a security analyst is required to merge similar clusters.
Alternatively, the maximum number of clusters to be estimated by X-Means can be set.
Subsequently, we perform an additional clustering step using context-less features to refine the clusters further.
For this purpose, we use the domain length, included  characters, used suffixes, Shannon entropy, and the number of contained English words.

In total, we identify 15 suspicious clusters including 45,189 samples (71.001\% of the investigated samples labeled as new DGA).
We present examples for each cluster in the Appendix~\ref{sec:appendix_new_clusters}.
We consider two of the clusters (Cluster 1 and 2) to be unknown DGAs based on their sample structure and sample size (a combined total of 35,814 samples).
Samples from the other clusters are often labeled as malicious or even DGA by VirusTotal~\footnote{https://www.virustotal.com}, but they are not included in DGArchive.
In addition, samples such as those found in Cluster 14 are included in phishing domain lists.

In general, it is possible to identify multiple clusters of unknown DGAs and malicious samples using only context-less features.
However, exploiting context information such as correlating queried domains labeled as belonging to new DGAs with the hosts that queried those domains could greatly improve the identification of unknown DGAs.
Further, the usage of contextual data could reveal additional clusters.
For instance, we reckon additional 9,817 samples to be DGA-generated but cannot make a definitive statement.
This cluster contains short and medium-length domains using a variety of 375 different TLDs.

%% file: content/adaptive_retraining.tex
\section{Adaptive Retraining}
\label{sec:adaptive_retraining}

In practice, it makes sense to adjust an existing DGA multiclass classifier as soon as a new DGA is discovered in order to be able to correctly assign samples to the newly found DGA.
This processes of adapting an already trained classifier by adding a new class to the classes learned by the classifier while trying to maintain the existing classification performance is called class-incremental learning.
Alternatively, a new classifier must be fully retrained, which could be more computationally expensive and require more training time.

Belouadah et al.~\cite{belouadah2021comprehensive} conducted a comprehensive study including multiple approaches to class-incremental learning and compared the approaches in six desirable properties.
In general, class-incremental learning is relevant in scenarios where memory and computing power are limited.
For instance, it might not be possible to store previously used training data due to storage space limitations.
In such a case, the classification accuracy of the old classes may decrease over time as new classes are gradually added to the classifier during new training epochs.
This problem is called catastrophic forgetting and was addressed by Castro et al.~\cite{castro2018end} by storing only a small representative set of training samples and using an adapted loss function that combines the cross-entropy loss to learn the new classes
with a distillation measure to retain the knowledge acquired from old classes.

Furthermore, class-incremental learning approaches often try to preserve the deep neural network structure as much as possible and integrate new information with only a minimal change in the classification layer~\cite{belouadah2021comprehensive}.
As a result, large deep learning networks do not have to be completely retrained, which saves valuable training time.
For instance, Lee et al.~\cite{lee2018simple} proposed a classification system that replaces the softmax classification layer with a classification based on the test sample's Mahalanobis distance to precomputed class-conditional Gaussian distributions.
There, the initially trained deep learning model is used as a feature extractor and is not retrained as soon as a new class is discovered.
Instead, a new class-conditional Gaussian distribution is extracted from the samples of the new class and added to the list of known distributions.

The disadvantage of such a method is that possible new features of the new classes are not learned by the deep learning classifier and therefore cannot be used for classification~\cite{castro2018end}.

Class-incremental learning can also be enabled for the examined approaches to the detection of new DGA families after a new class is discovered.
For the regex-based and Max-Softmax approaches, an additional output neuron needs to be added to the last classification layer.
For class-incremental learning, all weights except the last layer weights can be frozen, speeding up training because fewer weights need to be adjusted.
Finally, a new regex or class threshold must be computed to complete the training.
Similar to the class-incremental learning method proposed by Lee et al.~\cite{lee2018simple}, class-incremental learning can be enabled for the Family Vector approach.
This requires computing a new class cluster for the samples of the new class, which can be done without retraining the deep learning classifier used for feature extraction.
Likewise, for the One-Class Classifier approach, the deep neural network does not need to be retrained.
However, the one-class classifiers used require retraining, which accounts for most of the actual training time.

For the DGA detection use case, class-incremental learning speeds up the training time required to integrate new classes into an existing classifier.
However, it does not outweigh the fact that possible new features of new DGAs are not learned by the deep learning classifier and therefore cannot be used for classification.
Unlike other classification tasks such as image classification, the dataset containing a list of domain names is rather small and the deep learning model used is much less complex.
The ResNet model used for DGA multiclass classification contains 3.21 million parameters.
In comparison, the currently best performing deep learning model on the ImageNet benchmark contains 2.44 billion adjustable parameters~\cite{dai2021coatnet}.
Complete retraining is therefore comparatively cheap for DGA detection.
In fact, the complete retraining of the Regex - Error Detection approach, which mostly reduces to the retraining of the ResNet model, can be done in less than 15 minutes~\cite{drichel_analyzing_2020}.
Consequently, we propose to perform complete retraining as soon as enough samples belonging to unknown DGA families are detected.

%% file: content/conclusion.tex
\section{Conclusion}
\label{sec:conclusion}

In this work, we performed a comprehensive study using four different approaches in 15 different configurations to detect yet unknown DGAs.
This study includes the training of a total of 59,690 new DGA detection classifiers and an additional evaluation of 2,350 ensemble variants.
Our evaluation is based on a leave-one-group-out cross-validation using all 94 DGAs included in DGArchive.
Thereby, this evaluation scenario is particularly difficult and close to the real world, since the maximum number of known DGAs was used.
The examined Regex - Error Detection approach performed best and is able to detect several unknown DGAs with a high probability.
In detail, this approach is able to correctly attribute 14 out of 94 DGA classes with F1-scores above 80\%, 29 DGAs above 50\%, and 46 DGAs above 20\% while retaining state-of-the-art detection performance for known DGA families.
Using real-world data, we demonstrated the feasibility of this approach and detected 15 clusters of malicious samples and identified at least two clusters as unknown DGAs.
All examined classifiers operate without context and exclusively on a single domain name to be classified.
Therefore, these classifiers preserve privacy as much as possible and reduce potential concerns about providing classification as a service.
In a real-world deployment the classification capability of the proposed classifiers will not be exploited in isolation but rather in combination with contextual data.
The identification of a new DGA would not be based on the classification of a single domain, but rather on multiple classifications.
Correlating queried domains that are labeled as belonging to new DGAs with the hosts that queried those domains could significantly improve the correct detection of new DGAs.
In addition, we discussed different class-incremental learning strategies that adapt an existing classifier to newly discovered classes.
We concluded that for the DGA detection use case, a complete retraining makes the most sense whenever a new DGA is discovered.

%% file: content/appendix.tex
\section{New DGA Clusters}
\label{sec:appendix_new_clusters}

In Fig.~\ref{fig:new_dgas}, we show examples of each cluster of domain names labeled as new DGAs that we found during our analysis in Section~\ref{sec:separating_multiple_dgas}.

\section{Combining Approaches}
\label{sec:appendix}

In this section, we present an additional study trying to improve the classification performance by combining different approaches.

The Family Vector approach~(cf. Section~\ref{sec:fv}) can replace softmax classification of the conventional deep learning model for DGA multiclass classification by attributing an input sample to the class with the shortest class cluster center distance.
We evaluated this approach by performing a five-fold cross-validation using the dataset presented in Section~\ref{ref:dataset} and comparing it to the conventional softmax classification performed by the deep learning model.

In this evaluation, we observed that on average the softmax classification (F1-score of 80.813\%) outperforms the Family Vector classification (F1-score of 75.332\%).
However, upon examining the scores obtained for each individual DGA, we observed that the Family Vector classification scores higher in recall than the softmax classification for several DGAs.

\input{content/new_dgas.tex}

\begin{figure*}
	\centering
	\begin{subfigure}{1.0\columnwidth}
		\centering
		\includegraphics[width=\columnwidth]{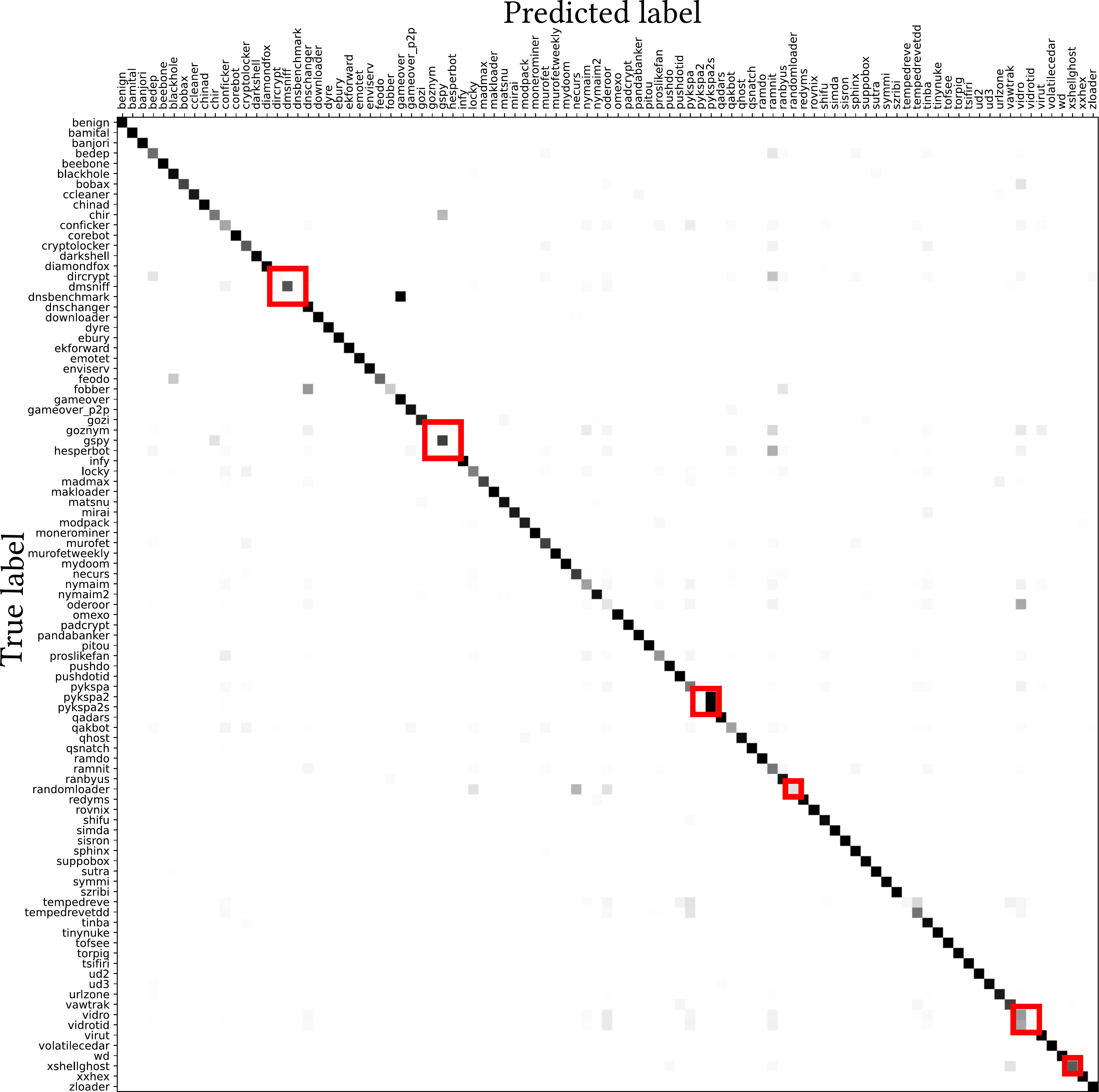}
		\caption{Softmax Classification}
		\label{fig:cm_bmc_softmax}
	\end{subfigure}
	\hfill
	\begin{subfigure}{1.0\columnwidth}
		\centering
		\includegraphics[width=\columnwidth]{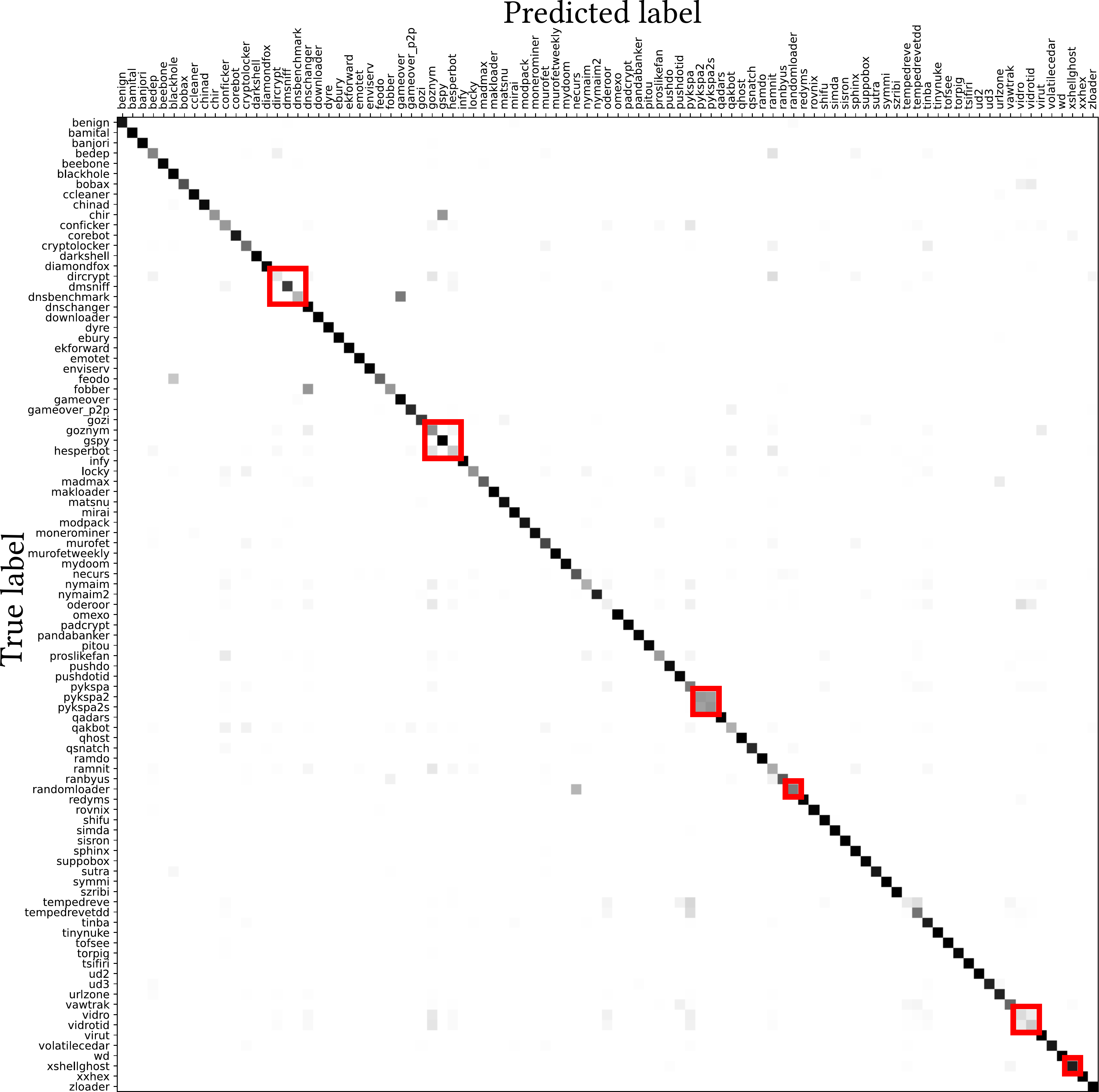}
		\caption{Family Vector Classification}
		\label{fig:cm_bmc_fv}
	\end{subfigure}
	\caption{Comparison of the confusion matrices of softmax and Family Vector classification, highlighting the key differences.}
	\label{fig:cm_bmc_comparision}
\end{figure*}

\input{tables/exploiting_results.tex}

In Fig.~\ref{fig:cm_bmc_comparision}, we visualize the confusion matrices for the softmax and Family Vector classification. 
For each classification approach, the confusion matrix displays the normalized proportion of samples with their true label on the y-axis that are attributed to the DGAs listed on the x-axis.
The portion of the samples is encoded in shades of gray, a pure black block corresponds to 100\% and a pure white block to 0\%.
A perfect classifier would yield a confusion matrix equal to the identity matrix.
We highlighted the key difference between the two classification approaches in red.

Especially for eight DGAs the Family Vector classification scores higher recall values.
These DGA classes are: \textit{dircrypt}, \textit{dmsniff}, \textit{dnsbenchmark}, \textit{goznym}, \textit{gspy}, \textit{hesperbot}, \textit{randomloader}, and \textit{xshellghost}.
Comparing the sample distribution in the dataset used (cf. Fig~\ref{fig:sample_distribution}), it can be seen that all of these classes are underrepresented and have significantly fewer samples than the maximum of 10,000 samples per class.
Therefore, for several weakly-represented classes, the Family Vector classification approach is better than softmax classification in separating samples from these classes.

We hypothesize that this is due to the fact that the softmax classification is biased towards well-represented classes, since more samples of the well-represented classes were presented to the classifier during training.
Earlier layers are less biased by this optimization because we assume that these layers act as feature extractors and are optimized independently of the class with all training samples and therefore are not biased in the same way as the softmax layer that does the final class attribution.
This allows Family Vector classification to better separate weakly-represented classes using early-layer features compared to softmax classification.

However, by biasing the softmax classification towards well-represented classes, higher precision values, and hence higher F1-scores, are obtainable because the classifier is more likely to be presented with a sample of a well-represented class.
This is also supported by our evaluation results.
The obtained precision value for Family Vector classification is 75.622\% and the one of softmax classification is 82.881\%.

In addition, Family Vector classification reduces confusion between two groups of related DGA families.
These groups are: \textit{pykspa2} -\textit{pykspa2s} and \textit{vidro}-\textit{vidrotid}.
We assume this is happening for the same reason as mentioned above.
Within the softmax classification confusion matrix, it can be seen that most samples from these two groups are attributed to a single DGA, namely to the DGA which is well-represented.
Comparing the confusion matrix of the Family Vector classification, it can be seen that the early-layer features allow for a partial separation of the related DGA families.

Based on these findings we try to combine different approaches to further improve classification results.
In detail, we try to (1) combine Family Vector classification with Regex Error - Detection, and (2) combine softmax classification with the Family Vector approach in order to increase the classification performance of weakly-represented classes.
For the latter, we tested multiple architectures with different approach prioritizations and thresholds based on sample support for classes within a used training dataset.

\begin{figure}
	\centering
	\includegraphics[width=0.572\columnwidth]{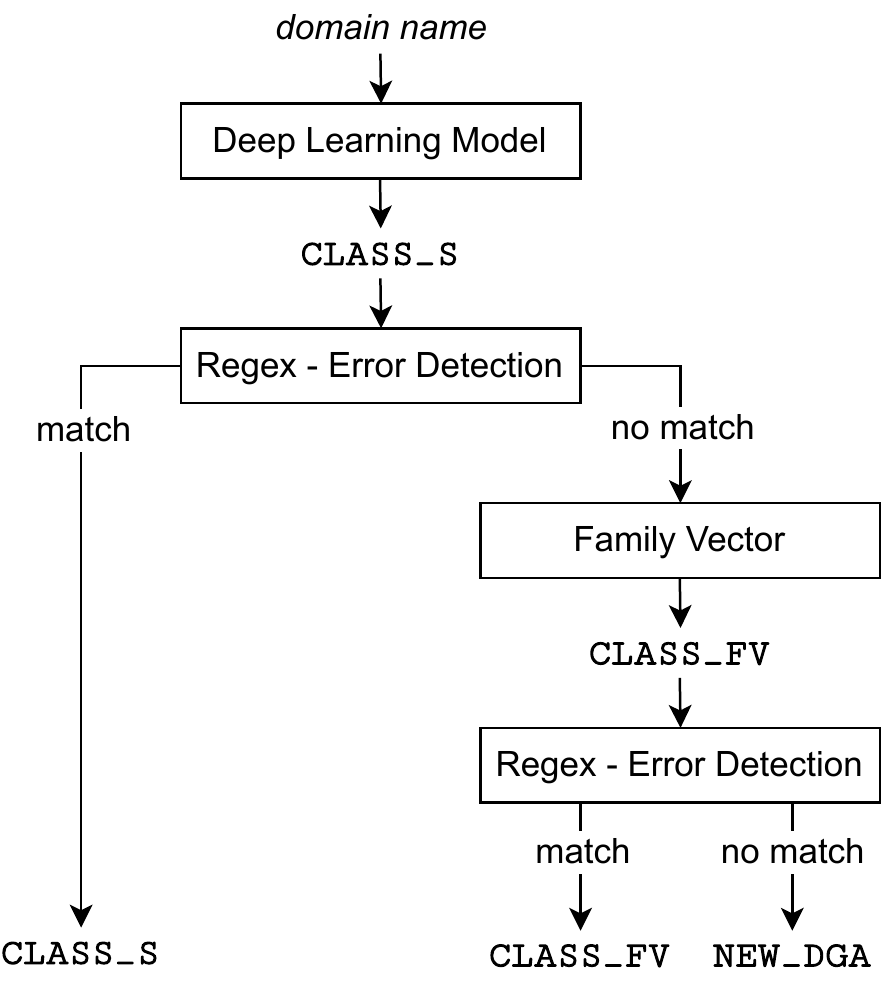}
	\caption{Structure of best performing combined classifier.}
	\label{fig:combination}
\end{figure}

We present the structure of the best performing combined classifier in Fig.~\ref{fig:combination}.
Class support thresholds were not beneficial and were therefore discarded.
The combined classifier first predicts a class using softmax classification and performs Regex - Error Detection.
If the predicted class \texttt{CLASS\textunderscore S} matches the regex it is returned.
Otherwise, Family Vector classification with subsequent Regex - Error Detection is performed.
If the predicted class \texttt{CLASS\textunderscore FV} matches the regex, it is returned, otherwise a sample is labeled as new DGA.

In Table~\ref{tab:combining_results}, we present the averaged five-fold cross-validated leave-one group-out evaluation results for the two combined approaches examined.

Combining the Family Vector approach with Regex Error - Detection improves the obtained evaluation scores compared to the plain Family Vector approach.
However, the achieved scores are still worse than the one of Regex Error - Detection using softmax classification.

The best combined classifier (cf. Fig.~\ref{fig:combination}) achieves an improvement in F1-score and LOC F1-score compared to Regex Error - Detection.
Comparing the F1-scores of the individual DGAs, no significant difference could be measured.
However, most classes are recognized with a slightly better probability.
While using other architectures for a combined classifier improved the correct detection of weakly-represented classes, the loss of precision was too great to justify.

In summary, the combined classifier performs slightly better than the best individual approach, Regex Error - Detection.
However, using two separate approaches is more computationally expensive.
It is questionable whether the slight increase in classification performance is worth the additional costs.

%% file: content/new_dgas.tex

\newcommand{\newagdsfontsize}{\fontsize{6.2}{6.2}\selectfont}

\begin{figure}
	\scriptsize
	\newagdsfontsize
	\centering
	\begin{subfigure}{0.495\linewidth}
		\begin{lstlisting}[columns=fullflexible,basicstyle=\ttfamily,belowskip=-1.5mm]
gmzkwhfcdaq.box					kumqnyvix.host
cvertlzgqlau.box				bdkuvgqrij.host
aywpxumhzkvqf.box			huxzkvnrwqg.host
		\end{lstlisting}
		\subcaption*{Cluster 1}
	\end{subfigure}

	\begin{subfigure}{1.0\linewidth}
	\begin{lstlisting}[columns=fullflexible,basicstyle=\ttfamily,belowskip=-1.5mm]
	j74dkz8l.2ney3vft.com
	q4y8be073g1va2s3huifg.ozazlqvrn5ixk6.com
	xet2dy6daffqpdxfm77lwe9v40b.f1znusp67s4k7gc1r0lzvkcj5al.com
	\end{lstlisting}
	\subcaption*{Cluster 2}
\end{subfigure}

	\begin{subfigure}{1.0\linewidth}
		\begin{lstlisting}[columns=fullflexible,basicstyle=\ttfamily,belowskip=-1.5mm]
	wvtkibfq.com.cn					vrfqdkhx.com.jp 				sjentlkh.com.tw				
	ftkhybgvq.com.cn				zguqbdlkh.com.jp				bdljgvtkh.com.tw			
	jguqbeovsj.com.cn			ichzguqbeo.com.jp			lkjentmnsh.com.tw		

		xybdkjfr.net.cn						mzxfjtptnk.org.ir
		zdmouqadk.net.cn					vkmecvuqthj.org.ir
		pzfsibdlkj.net.cn 			bwszdbulvywk.org.ir
		\end{lstlisting}
	\subcaption*{Cluster 3}
	\end{subfigure}

\begin{subfigure}{1.0\linewidth}
	\begin{lstlisting}[columns=fullflexible,basicstyle=\ttfamily,belowskip=-1.5mm]
	downloads.76da979ed84e771184d2bef6558822c01d8d755d.com
	downloads.76da979ed84e771184d2bef6558822c01d8d755d.info
	downloads.fd9fc45e124ee95dd89dbddf6409282a0c8eb24c.com
	\end{lstlisting}
	\subcaption*{Cluster 4}
\end{subfigure}

\begin{subfigure}{0.495\linewidth}
	\begin{lstlisting}[columns=fullflexible,basicstyle=\ttfamily,belowskip=-1.5mm]
	2f782a4fa1.pw
	67df4075e33e.us
	279e0ab651512a836b7f.com
	\end{lstlisting}
	\subcaption*{Cluster 5}
\end{subfigure}
\begin{subfigure}{0.495\linewidth}
	\begin{lstlisting}[columns=fullflexible,basicstyle=\ttfamily,belowskip=-1.5mm]
	1xredwlz.host
	www.1xxpers130.mobi
	www.1xredhkcb.world
	\end{lstlisting}
	\subcaption*{Cluster 6}
\end{subfigure}

\begin{subfigure}{0.495\linewidth}
	\begin{lstlisting}[columns=fullflexible,basicstyle=\ttfamily,belowskip=-1.5mm]
a.41a9d21468d3f181278201d6b84e8d0a.com
a.7c43c18ac8544f55141109e369cca87c.com
a.2b4764540443253ac540b82440a854ad.com
	\end{lstlisting}
	\subcaption*{Cluster 7}
\end{subfigure}
\begin{subfigure}{0.495\linewidth}
	\begin{lstlisting}[columns=fullflexible,basicstyle=\ttfamily,belowskip=-1.5mm]
	     afa-02.com
	     afa-06.com
	     afa-10.com
	\end{lstlisting}
	\subcaption*{Cluster 8}
\end{subfigure}

\begin{subfigure}{0.495\linewidth}
	\begin{lstlisting}[columns=fullflexible,basicstyle=\ttfamily,belowskip=-1.5mm]
5rbvtfbij44lgr871gm820sl3a9ld03n.net
db7mrfqq3le6j23a19cvvmo6cqlgk49f.net
2qljtiqfms5h3s0n7vl98n51okhn3e6a.net
	\end{lstlisting}
	\subcaption*{Cluster 9}
\end{subfigure}
\begin{subfigure}{0.495\linewidth}
	\begin{lstlisting}[columns=fullflexible,basicstyle=\ttfamily,belowskip=-1.5mm]
	  s.vmjwlyto.ru
	  s.zkyibuwgp.com
	  s.lqpawkujw.com
	\end{lstlisting}
	\subcaption*{Cluster 10}
\end{subfigure}

\begin{subfigure}{0.495\linewidth}
	\begin{lstlisting}[columns=fullflexible,basicstyle=\ttfamily,belowskip=-1.5mm]
	    www.hahb.bid
	    www.hahm.bid
	    www.hahz.bid
	\end{lstlisting}
	\subcaption*{Cluster 11}
\end{subfigure}
\begin{subfigure}{0.495\linewidth}
	\begin{lstlisting}[columns=fullflexible,basicstyle=\ttfamily,belowskip=-1.5mm]
	     aab13.online
	     abbb2.online
	     abbe1.online
	\end{lstlisting}
	\subcaption*{Cluster 12}
\end{subfigure}

\begin{subfigure}{0.495\linewidth}
	\begin{lstlisting}[columns=fullflexible,basicstyle=\ttfamily,belowskip=-1.5mm]
    a2.security-service1.info
    p1.security-service16.info
    ga3.security-service31.info
	\end{lstlisting}
	\subcaption*{Cluster 13}
\end{subfigure}
\begin{subfigure}{0.495\linewidth}
	\begin{lstlisting}[columns=fullflexible,basicstyle=\ttfamily,belowskip=-1.5mm]
    trustedpagesvalidation02.ga
    trustedpagesvalidation03.ml
    trustedpagesvalidation03.gq
	\end{lstlisting}
	\subcaption*{Cluster 14}
\end{subfigure}

\begin{subfigure}{0.495\linewidth}
	\begin{lstlisting}[columns=fullflexible,basicstyle=\ttfamily,belowskip=-1.5mm]
	maren740.xyz
	lc.ellena485.xyz
	qcretzdzvv.osborn283.xyz
	\end{lstlisting}
	\subcaption*{Cluster 15}
\end{subfigure}

	\caption{Suspicious clusters of new DGA labeled domains.}
	\label{fig:new_dgas}
\end{figure}

%% file: tables/exploiting_results.tex

\renewcommand{\arraystretch}{0.5}

\begin{table*}[!t]
	\caption{Evaluation results for combined approaches.}
	\label{tab:combining_results}
	\centering
	\resizebox{0.916\linewidth}{!}{
		\begin{tabular}{l|ccc|ccc}
			\hline \Tstrut
			\multirow{2}{*}{\textbf{Approach}} & \multirow{2}{*}{\textbf{F1-Score}} & \multirow{2}{*}{\textbf{Precision}} & \multirow{2}{*}{\textbf{Recall}} & \textbf{LOC} & \textbf{LOC} & \textbf{LOC} \\
			& & & & \textbf{F1-Score} & \textbf{Precision} & \textbf{Recall} \\
			\hline \Tstrut
			Family Vector - Mean Threshold + Regex - Error Detection & 0.79237 & 0.80157 & 0.82189 & 0.27797 & 0.26976 & \textbf{0.39026} \\
			\hline \Tstrut
			Best Combined Classifier (cf. Fig.~\ref{fig:combination}) & \textbf{0.80355} & \textbf{0.82080} & \textbf{0.80958} & \textbf{0.32701} & \textbf{0.46595} & 0.31950 \\
			\hline
		\end{tabular}
	}
\end{table*}

\renewcommand{\arraystretch}{0.947}